\documentclass[aps,prc,twocolumn,floatfix,nofootinbib,preprintnumbers,superscriptaddress,longbibliography]{revtex4-1}

\usepackage{epsfig}
\usepackage{graphicx}
\usepackage{stmaryrd}
\usepackage{amssymb,tabularx,dcolumn}
\usepackage{supertabular,ltxtable}
\usepackage{longtable}
\usepackage{amsmath}
\usepackage{amstext} 
\usepackage{float}
\usepackage{color}
\usepackage{bm}
\usepackage{CJKutf8}
\usepackage{hyperref}
\usepackage{mathtools}
\usepackage{multirow}
\usepackage{makecell}
\usepackage[utf8]{inputenc}
\usepackage{tabu}
\usepackage{braket}

\makeatletter
\def\@bibdataout@aps{%
\immediate\write\@bibdataout{%
@CONTROL{%
apsrev41Control%
\longbibliography@sw{%
    ,author="08",editor="1",pages="1",title="0",year="1"%
    }{%
    ,author="08",editor="1",pages="1",title="",year="1"%
    }%
  }%
}%
\if@filesw \immediate \write \@auxout {\string \citation {apsrev41Control}}\fi 
}
\makeatother

\renewcommand{\vec}[1]{\mbox{\boldmath $#1$}}

\renewcommand{\vec}[1]{\mbox{\boldmath $#1$}}

\usepackage{hyperref}

\definecolor{star}{RGB}{31,119,180}

\begin{document}
\begin{CJK*}{UTF8}{gbsn}
\date{today}

\title{Gamow Shell Model description of  Li isotopes and their mirror partners}

\author{X. Mao (毛兴泽)}
\affiliation{National Superconducting Cyclotron Laboratory, Michigan State University, East Lansing, Michigan 48824, USA}
\affiliation{Department of Physics and Astronomy, Michigan State University, East Lansing, Michigan 48824, USA}

\author{J. Rotureau}
\affiliation{National Superconducting Cyclotron Laboratory, Michigan State University, East Lansing, Michigan 48824, USA}

\author{W. Nazarewicz}
\affiliation{Facility for Rare Isotope Beams, Michigan State University, East Lansing, Michigan 48824, USA}
\affiliation{Department of Physics and Astronomy, Michigan State University, East Lansing, Michigan 48824, USA}

\author{N. Michel}
\affiliation{Institute of Modern Physics, Chinese Academy of Sciences, Lanzhou, Gansu 730000, PR China}

\author{R.M. Id Betan}
\affiliation{Physics Institute of Rosario (CONICET), Rosario, Argentina}
\affiliation{Department of Physics FCEIA (UNR), Rosario, Argentina}

\author{Y. Jaganathen}
\affiliation{Institute of Nuclear Physics, Polish Academy of Sciences, PL-31342 
Kraków, Poland}
\date{\today}

\begin{abstract}
  \textbf{Background:}
Weakly bound and unbound nuclei close to  particle drip lines are laboratories of new nuclear structure physics at the extremes of neutron/proton excess. 
The comprehensive  description of  these systems  
requires an open quantum system framework that is capable of treating resonant and nonresonant many-body states on equal footing.

\textbf{Purpose:} 
In this work, we  develop  the complex-energy configuration interaction  approach to describe binding energies and spectra  of  selected $5\le A \le 11$ nuclei.

\textbf{Method:}
We employ the complex-energy Gamow shell model (GSM) assuming a  rigid $^4$He core.
The effective Hamiltonian, consisting of a core-nucleon Woods-Saxon potential and  a simplified version of the Furutani-Horiuchi-Tamagaki interaction with the mass-dependent  scaling, is optimized in the $sp$ space. 
To diagonalize 
the Hamiltonian matrix, we employ the Davidson method and the Density Matrix Renormalization Group technique.

\textbf{Results:} 
Our optimized  GSM Hamiltonian offers a good reproduction of binding energies and spectra with the  root-mean-square  (rms)  deviation from experiment of 160\,keV.
Since the model  performs well when used to predict known excitations
that have not been  included in the fit, it can serve as a reliable tool to describe  poorly known  states. A case in point is our prediction for the pair 
of  unbound mirror nuclei  $^{10}$Li-$^{10}$N in which a huge Thomas-Ehrman shift dramatically alters the pattern of  low-energy excitations.

\textbf{Conclusion:}
The new model will enable comprehensive studies of structure and reactions
aspects of light drip-line nuclei.
\end{abstract}

\maketitle
\end{CJK*}

\section{Introduction}{\label{introduction}}

With progress in radioactive beam
experimentation and many impressive advances  in the microscopic
nuclear theory, light nuclei provide an excellent 
ground for testing both nuclear
interactions and many-body approaches. Of particular interest are weakly bound and unbound nuclear systems with extreme neutron-to-proton imbalance, whose structure is profoundly affected by the coupling to the continuum of decay and reaction channels \cite{Jonson2004,AlKhalili2003,Dobaczewski2007,Forssen2013}. 

The important challenge for the field of low-energy nuclear theory  is to
unify nuclear
bound states with resonances and scattering continuum within one consistent framework
\cite{Johnson2019}. There are many open questions that can be answered by studying drip-line systems \cite{Michel2010}: 
What can be said about properties of weakly bound or unbound many-body systems close to the reaction threshold? Do their properties depend on any particular realization of the Hamiltonian? Which nuclear properties are  impacted by the coupling to the  continuum of scattering and decaying states? Theoretically, a coherent description of the interplay between  bound and unbound states  in the many-body  system requires an  open quantum system   formulation. In this context, this area of research is truly interdisciplinary. Indeed, open quantum systems are  studied in various fields of physics: nuclear physics, atomic and molecular physics, nanoscience, quantum optics, etc. In spite of their differences, such systems  often display universal properties that are common to all weakly bound or unbound systems close to the reaction/decay threshold.

Impressive progress has been achieved in describing weakly bound and unbound nuclei using $A$-body methods rooted in realistic inter-nucleon interactions
 \cite{Hagen14,Hagen2016,Navratil16,Quaglioni2018,Freer2018}. Examples include microscopic computations of $^{11}$Be \cite{Calci2016,Bonaccorso_2019_2067}, 
 $^7$He \cite{Baroni_2013a,Baroni_2013b,Mazur2019}, and
$^{9}$He \cite{Vorabbi2018}. 

On a more phenomenological level, configuration integration techniques, based on the concept of valence nucleons coupled to an inert core have reached a high level of sophistication.
Approaches such as the real-energy continuum shell model \cite{volya06_94,Volya2014} and shell model embedded in the continuum  \cite{bennaceur98_99, Okolowicz_2003_2042,rotureau06_43,Okolowicz2020} have been applied to systems near particle-emission threshold with  one/two  nucleons allowed in the continuum space. Another powerful tool is
the complex-energy Gamow Shell Model (GSM) \cite{Michel2003,betan04_37,michel09_2}, an extension of the interacting shell
model to the treatment of open quantum systems. GSM has been successfully used 
to describe structural and reaction properties of exotic nuclei
(see Refs. \cite{fossez17_1916,Michel_2019_2063,Mercenne_2019_2064,Jones_2017_1993,Yannen2017_1988} for recent representative applications).

This study can be viewed as a continuation of previous work on the development of a quantitative GSM description of 
light nuclei using  a $^4$He-nucleon potential and finite-range  interaction between valence nucleons. In the first paper \cite{Yannen2017_1988}, where calculations were carried out in the $spdf$ model space, the core-nucleon potential was optimized to nucleon-$^4$He phase shifts. By means of the principal-component analysis, it was concluded that a very reasonable description of energies of $6\le A \le 9$ nuclei 
(with the root-mean-square (rms) deviation from experiment
of 250 keV)  could be achieved with only four interaction
parameters. In the follow-up study  \cite{Fossez_2018_1994}, where calculations were performed in  the $spd$
space, experimental energies and widths of $^{5-8}$He  could be reproduced within tens of keV precision by
adjusting only one parameter (the strength of spin-singlet central neutron-neutron term). 
In this work, we  use the  GSM model to describe binding energies and spectra  of  $5\le A \le 11$ nuclei in the $sp$ space by carrying out simultaneous optimization of the core-nucleon potential and the valence two-body interaction with the mass-dependent interaction scaling to effectively account for the missing  three-body forces. 
We show that with the appreciable reduction of the parameter space (four strengths of the core-nucleon  potential and four parameters
of the two-body interaction), a very reasonable agreement with experimental energies is obtained.

 Predictions were also made for the  particle-unstable nuclei $^{10}$Li, $^{10}$N, and $^{11}$O, which are excellent laboratories of open quantum system physics. In particular,  a spectacularly strong Thomas-Ehrman effect
in the  $^{10}$N-$^{10}$Li mirror pair is predicted.

This paper is organized as follows.
The theoretical model is outlined in Sec.~\ref{theoretical_model}, which contains a short
overview of GSM, description of the GSM Hamiltonian, and the  optimization protocol.
Results are presented in Sec.~\ref{results}, with the  optimization results discussed first, followed by  predictions for lithium isotopes and their mirror partners.
Finally, Sec.~\ref{conclusion}  presents conclusions
and perspectives for future studies.


\section{Theoretical Model} \label{theoretical_model}

\subsection{Gamow Shell Model}\label{GSM}
Here we briefly recall the GSM formalism.
In this work, we describe the lithium isotopes and their mirror partners in terms of
valence nucleons coupled to the   $^4$He core.
This picture is justified by the fact that the $^4$He nucleus is a tightly bound system with   the first excited state  located
20.21 MeV above the ground state (g.s.)  \cite{ensdf}.

The GSM Hamiltonian can be written as
\begin{equation}{\label{eq:H}}
H=\sum_i^{N_{\rm val}}\left[\frac{\vec{p}_i^2}{2\mu_i} +U_{\rm c}(i)\right] + \sum_{i=1,j>i}^{N_{\rm val}}\left[ V_{i,j} + \frac{\vec{p}_i\vec{p}_j}{M_{\rm c}} \right],
\end{equation}
where $N_{\rm val}$ denotes the number of valence nucleons, $\mu_i$ and $M_{\rm c}$ are the reduced mass of the nucleon and the mass of the core, respectively,
 $U_{\rm c}$ is the core-nucleon potential, and  $V_{i,j}$ is the interaction between valence nucleons. The Hamiltonian (\ref{eq:H})  is written in the cluster orbital shell model coordinates \cite{suzuki88_595} defined with respect to the center of mass of the core.

The GSM  Hamiltonian is diagonalized in the Berggren basis \cite{berggren68_32},
which allows to consistently treat bound, resonance, and scattering states.
In the complex-momentum space, the Berggren basis obeys the closure relation for each partial wave $(\ell,j)$:
\begin{equation} {\label{eq:Berggren}}
\sum_{n=b,d} \ket{\tilde{u}_n} \bra{u_n} + \int_{\mathcal{L}^+} \ket{\tilde{u} (k)} \bra{u(k)} dk =1,
\end{equation}
where $b$ and $d$ stand for the bound states and selected decaying resonant states, respectively, and the contour $\mathcal{L}^+$  representing the non-resonant scattering states is located in the fourth quadrant of the complex $k$-plane. 
The specific shape of  $\mathcal{L}^+$  is not important 
as long as all resonant states between the real axis and the contour  $\mathcal{L}^+$ are included.
In practical applications, the contour is discretized for each $(\ell,j)$, which results in a finite number of single-particle (s.p.) states. From this discretized set of shells
one constructs Slater determinants, which form a many-body basis within which the GSM Hamiltonian is diagonalized. 
Due to the inclusion of resonances and complex-momentum scattering states, the Hamiltonian representation  in the Berggren basis is complex symmetric \cite{michel09_2}.

As in any configuration interaction approach, the dimension of the Hamiltonian matrix grows quickly with the number of active particles.
In the context of the GSM,  it increases more quickly than in the conventional shell model due to the presence of discretized scattering states.
To this end, we truncate  the model space
by working with natural orbitals which provide an optimized set of s.p. states \cite{brillouin33,shin16_1860,Yannen2017_1988}.

The natural orbitals are first computed in a  truncated space where  few valence particles are allowed
to occupy  continuum shells. A truncation is then performed on the s.p. basis by keeping
only natural orbitals for which the modulus of the  occupation number is greater
than a certain (small)  value. Finally, a new set of Slater determinants is constructed, for which also a truncation on the number of particles in the continuum
is enforced, and the numerical diagonalization is performed  using the Davidson method \cite{Jacobi_Davidson}.

To check the accuracy of this truncation procedure in the case of the largest systems,
a supplementary computation was also performed using the
Density Matrix Renormalization Group (DMRG) \cite{rotureau06_15,rotureau09_140} method.
The DMRG allows performing calculations without  the s.p. particle basis truncation and without restrictions on the number  of particles
in the continuum. In this approach, the  many-body Schr\"odinger equation is solved iteratively
in tractable truncated spaces, which are gradually increased until the numerical convergence is reached.
We have checked that, in all cases discussed in this work, the GSM results are in good agreement with those of DMRG (see more discussion in Sec.~\ref{computation}).

\subsection{Gamow Shell Model Hamiltonian}\label{Interaction}

The core-nucleon potential  is taken as a Woods-Saxon (WS) field, with a central and spin-orbit terms, and the Coulomb field for protons:
\begin{equation}
U_{\text{c}}(r)=V_0 f(r)-4V_{\ell s} \frac{1}{r}\frac{df(r)}{dr} \boldsymbol{\ell} \cdot \boldsymbol{s} + U_{\text{Coul}}(r),
\end{equation}
where $f(r)=-(1+\text{exp}[(r-R_0)/a])^{-1}$.
The WS radius $R_0$ and diffuseness $a$  were taken from 
Ref.~\cite{Yannen2017_1988}: $R_0(n)=2.15$\,fm, $R_0(p)=2.06$\,fm, $a(n)=0.63$\,fm,
and $a(p)=0.64$\,fm.
The Coulomb potential is  generated by a spherical Gaussian charge distribution with   radius $R_{ch}=1.681$\,fm \cite{Ingo_2008_1992}.

Following Ref.~\cite{Yannen2017_1988}, the interaction between valence nucleons is  a sum of central, spin-orbit,  tensor, and  Coulomb terms:
\begin{equation}
V = V_{\rm c} + V_{\rm LS} +V_{\rm T} +V_{\rm Coul}.
\end{equation}
The central, spin-orbit and tensor interactions are  constructed based on 
the finite-range Furutani-Horiuchi-Tamagaki (FHT) force \cite{furutani78_1012,furutani79_1013,Yannen2017_1988}.
For each term, the radial form factor is represented by
a sum of three Gaussians with different widths representing the short, intermediate and long ranges of the nucleon-nucleon interaction.
This interaction has been used successfully to describe  structure and reactions involving light nuclei \cite{fossez17_1927,Yannen2017_1988,fossez16_1793,Jones_2017_1993,Fossez_2018_1994,Michel_2019_2063,Mercenne_2019_2064}.

In order to be applied in the present GSM formalism, the interaction is rewritten in terms of the spin-isospin projectors $\Pi_{ST}$ \cite{RingSchuck}:
\begin{equation}\label{twobody}
\begin{aligned}
V_{\rm c}(r) & = V_c^{11}f_c^{11}(r) \Pi_{11} + V_c^{10}f_c^{10}(r) \Pi_{10}  \\
& + V_c^{00}f_c^{00}(r) \Pi_{00} + V_c^{01}f_c^{01}(r) \Pi_{01},  
\\
V_{\rm LS} &= (\boldsymbol{L}\cdot \boldsymbol{S})\,V_{LS}^{11}f_{LS}^{11}(r) \Pi_{11},\\
V_{\rm T}(r)& =S_{ij}\left[V_T^{11}f_T^{11}(r) \Pi_{11} + V_T^{10}f_T^{10}(r) \Pi_{10}\right],
\end{aligned}
\end{equation}
where $r\equiv r_{ij}$ stands for the  distance between the nucleons $i$ and $j$, 
 $\hat{r}=\vec{r}_{ij}/r_{ij}$, 
$\vec{L}$ is the relative orbital angular momentum,  $\vec{S}=(\vec{\sigma}_i+\vec{\sigma}_j) / 2$,  and  $S_{ij}=3 (\vec{\sigma}_i \cdot \hat{r}) (\vec{\sigma}_j \cdot \hat{r})   - \vec{\sigma}_i \cdot \vec{\sigma}_j$. 
The interaction (\ref{twobody}) is characterized by
 the seven interaction strengths in spin-isospin channels,
$V_c^{11}$, $V_c^{10}$, $V_c^{00}$, $V_c^{01}$, $V_{LS}^{11}$, $V_{T}^{11}$, and   $V_{T}^{10}$.

In Ref.~\cite{Yannen2017_1988}, the FHT interaction was used in the GSM description of bound and unbound nuclei  with $A \leq 9$.
While a good energy reproduction  was achieved, the systematic statistical study of the parameters carried out in Ref.~\cite{Yannen2017_1988}
demonstrated that some of the terms in the FHT interaction  were sloppy, i.e., not well constrained.

In this study, we use a simplified version of the FHT interaction
where we  consider the central $V_c^{10}, V_c^{01}$, and  tensor  $V_T^{10}$ terms.
This choice is not only informed by the previous statistical work \cite{Yannen2017_1988} 
but also justified by Effective Field Theory (EFT) arguments \cite{Ordonez_1992_2078,Bedaque02,Bedaque03,stetcu10,Capel2018}.
Indeed, in the EFT expansion of the bare nucleon-nucleon interaction, these three terms 
appear at leading order, whereas the other terms present in the original FHT interaction  correspond to  higher orders of EFT.
However, we have observed that adding the central term $V_c^{00}$
improves the overall description of the nuclei considered in this work and hence
 we  have also included it in  $V_{i,j}$.
We want to mention here that a similar approach was employed in Ref.~\cite{Fossez_2018_1994} to construct an effective neutron-neutron interaction for the description of the helium isotopic chain  in the Berggren basis. In that case, using only the central term  $V_c^{01}$, a good reproduction of weakly-bound and unbound states in helium nuclei was achieved.

As it is customary in shell model studies \cite{Alex2006_1989,Huth2018_1990}, 
a mass-dependent interaction-scaling factor of the form $(6/A)^{\alpha}$ is introduced 
to effectively account for the missing three-body forces \cite{zuker03_1026,Stroberg_2019_2046}.
We found that the value $\alpha=1/3$ gives a very reasonable description  of experimental energies. Finally, the  Coulomb interaction between valence protons is treated  by incorporating its long-range part into the basis potential 
and expanding the short-range two-body component in a truncated basis of HO states \cite{hagen06_14,michel10_3}.

\subsection{Interaction Optimization Protocol}\label{Interaction_optimization}

Our interaction optimization protocol strictly follows that of
Ref.~\cite{Yannen2017_1988}. In short,  we minimize the $\chi^2$ penalty function:
\begin{equation}\label{eq:chi}
\chi^2(\vec{p})=\sum_{i=1}^{N_d} \bigg( \frac{\mathcal{O}_i(\vec{p})- \mathcal{O}_i^{\mbox{\scriptsize{exp}}}}{\delta \mathcal{O}_i } \bigg)^2
\end{equation}
where $\vec{p}$ is the vector of parameters used, $N_d$ is the number of observables,
$\mathcal{O}_i(\vec{p})$ are the calculated  observables,  $\mathcal{O}_i^{\mbox{\scriptsize{exp}}}$ are experimental values, and $\delta \mathcal{O}_i$ are the adopted  errors that have been obtained from the $\chi^2$ normalization  \cite{Birge_1932_1996,dobaczewski14_1134}.

The minimization of $\chi^2$ is done using the Gauss-Newton method. Since
the GSM Hamiltonian  is linear in the strength parameters,  the Jacobian matrix at the minimum $\vec{p}_0$,
\begin{equation}\label{Jac}
J_{i\alpha}=\frac{1}{\delta \mathcal{O}_i} \left.\frac{\partial \mathcal{O}_i}{\partial p_{\alpha}} \right|_{\vec{p}_0},
\end{equation}
can be calculated exactly using the Hellmann-Feynman theorem \cite{feynman39_1069}.
The covariance matrix $\mathcal{C}$ can be expressed in terms of $J$:
\begin{equation}
\mathcal{C}\simeq (J^TJ)^{-1}
\end{equation}
In the situation where the Jacobian matrix is non-invertible or has a very small determinant, the Gauss-Newton method becomes unstable.
This typically happens when a parameter is sloppy, i.e.,  not well constrained by observables.
In order to stabilize the calculation, the matrix inversion is replaced by its pseudo-inverse, derived from the singular value decomposition (SVD) of the Jacobian matrix \cite{Yannen2017_1988}.

The uncertainties on parameters and predicted observables can be computed with the help
of the covariance matrix $\mathcal{C}$.
For more details, the reader is referred to Ref.~\cite{Yannen2017_1988}.

The four strengths of the WS potential and four parameters of the two-body interaction are simultaneously optimized to reproduce
15 energy levels in lithium isotopes and their mirror partners given in  Table~\ref{Table:Opt_states}. 
\begin{table}[htb]
\caption{\label{Table:Opt_states} Energy levels used in the GSM Hamiltonian optimization.
The energies are given with respect to the $^4$He g.s.
The experimental values $E_{\text{exp}}$ are taken from \cite{ensdf}. They are compared to the GSM values $E_{\text{GSM}}$.
}
\begin{ruledtabular}
\begin{tabular}{ c  c  c  c  }
	Nucleus	 &	State&	$E_{\text{exp}}$\,(MeV)		& 	$E_{\text{GSM}}$\,(MeV)		\\
\hline  \\[-6pt]
\multirow{2}{*}{$^6$Li} 	&	1$^+$		& $-$3.70	& $-$3.72					\\
		&	0$^+$	& $-$0.14	&$-$0.10						\\[3pt]
\multirow{2}{*}{$^7$Li} 	&	3/2$^-$ 		& $-$10.95&$-$11.02						\\
		&	1/2$^-$		& $-$10.47&$-$10.14						\\[3pt]
\multirow{2}{*}{$^8$Li} 	&	2$^+$ 	& $-$12.98		&$-$13.14					\\
		&	1$^+$		& $-$12.00 &$-$11.93						\\[3pt]
\multirow{2}{*}{$^9$Li} 	&	3/2$^-$ 	& $-$17.05		&$-$16.90					\\
		&	1/2$^-$ 	& $-$14.35		&$-$14.50					\\[3pt]
$^{11}$Li 	&	3/2$^-$ 	& $-$17.41	&$-$17.48				  		\\[3pt]
\multirow{2}{*}{$^7$Be} 	&	3/2$^-$ 	& $-$9.30		&$-$9.36					\\
		&	1/2$^-$ 		& $-$8.88		&$-$8.53				\\[3pt]
\multirow{2}{*}{$^8$B} 	&	2$^+$ 	& $-$9.44			&$-$9.60				\\
		&	1$^+$ 	& $-$8.67		&$-$8.50					\\[3pt]
\multirow{2}{*}{$^9$C} 	&	3/2$^-$ 	& $-$10.74	&$-$10.85						\\
		&	1/2$^-$ 	& $-$8.52		&$-$8.59 		
\end{tabular}
\end{ruledtabular}
\end{table}

The calculations are performed in a model space which includes $s_{1/2}$, $p_{3/2}$, and $p_{1/2}$ partial waves for both protons and neutrons.
Since the optimization  involves energies only, for the sake of speeding-up the optimization and for better stability, we used a deeper WS potential to generate the basis, in which the $0p_{3/2}$ and $0p_{1/2}$ poles are bound.
A real contour was then used to describe the non-resonant continuum space.
The contour $\mathcal{L}^+$, independent of interaction parameters,  was divided into 3 segments: $[0,k_{\rm peak}], [k_{\rm peak},k_{\rm mid}]$, and  $[k_{\rm mid},k_{\rm max}]$,  with the values $k_{\rm peak}= 0.25$\,fm$^{-1}$,  $k_{\rm mid}= 0.5$\,fm$^{-1}$, and the cutoff momentum $k_{\rm max}= 4$\,fm$^{-1}$.
Discretizing each segment with 10 points using the Gauss-Legendre quadrature guarantees the convergence of results.

To calculate resonance's width, one has to generate a  basis  based on a shallower basis-generating WS potential, in which the  $0p_{3/2}$ and $0p_{1/2}$ poles are decaying resonances. In this case,
a complex contour defined by a complex value of $k_{\rm peak}$  is employed.
It is to be noted that calculation of the width is more demanding than that of energy. 
A higher discretization with 20 points for each segment was used for this purpose.
Due to the Coulomb repulsion, the mean field used to generate the s.p. basis for proton rich nuclei varies with proton number.
The contour is adjusted separately for each system to assure that the Berggren completeness relation is met.
To ensure the numerical stability, the chosen contour should neither lie too close to the Gamow poles nor lie too far from the real-$k$ axis.
In this work, $k_{\rm peak}$ is chosen to lie slightly below the position of the $0p_{3/2}$, $0p_{1/2}$ poles, but with the imaginary part greater than $-0.2$\,fm$^{-1}$.
The calculations were repeated with several slightly different values of $k_{\rm peak}$ to assure the full convergence.

\subsection{Computational Details}\label{computation}

In this study, we used a newly developed GSM code that is based on the two-dimensional partitioning of the Hamiltonian matrix \cite{Michel_2020_2076}.
First, we computed natural orbitals from a  calculation with at most two particles in the continuum space. The s.p. basis was further truncated 
by keeping the natural orbitals with occupations  greater than $10^{-6}$.
The GSM problem was then solved in a model space with at most four particles in the continuum shells.
We checked the accuracy of this truncation by performing  full DMRG calculations for the  systems with $A=9-11$.

The DMRG allows the computations of energies without truncation in the s.p. basis and without restriction on the number  of particles in the continuum.
In the first stage of the DMRG procedure, the set of shells is split into two subsets $H$ and $P$: the pole subspace $H$ consists of  the Gamow poles considered   (for instance, in the DMRG computations of the  $^{9}$Li g.s., $H$ contains the $0p_{3/2}$
and $0p_{1/2}$ Gamow states) and the remaining shells form the subspace $P$. The resolution of the  Schr\"odinger equation is then performed in an increasing set of shells, by gradually including the shells of $P$, one at a time.
After having considered a given shell of $P$, the model space is truncated by keeping $N_{\rm kept}$ many-body states that  correspond to the
eigenstates  of the density matrix with the largest eigenvalues $w_i$ (in modulus). The number of states kept is defined by the control parameter $\epsilon$ so that the condition $|1-\Re(\sum^{N_{\rm kept}}_{i=1}) w_i| < \epsilon$
is fullfilled. The first DMRG stage  ends when all shells in $P$ have been included. At that point, natural orbitals are computed
and  new subsets $H$ and $P$ are defined. The new subset $H$ contains $N_H$ natural orbitals. The calculation continues in a similar fashion, by adding shells from the new subset $P$, one by one, until all shells have been considered, 
and then a new set of natural orbitals is computed. $N_H$ is increased and $\epsilon$ decreased, until convergence (few keV), is achieved.
For instance, in the case of the $^{9,11}$Li g.s.,
computations were carried out by increasing $N_H$ up to 12 and $\epsilon$ was decreased down to $5\times 10^{-9}$ (a typical DMRG accuracy \cite{Fossez_94_Mg}).
 For both nuclei, the GSM energies turned out to be  less than 10 keV above the DMRG results.   For more details about our DMRG implementation, see \cite{rotureau06_15,rotureau09_140,shin16_1860}. 
%
\section{Results} \label{results}
\subsection{Optimized Interaction}\label{optimize_interaction}

\begin{figure*}[!htb]
\includegraphics[width=0.7\linewidth]{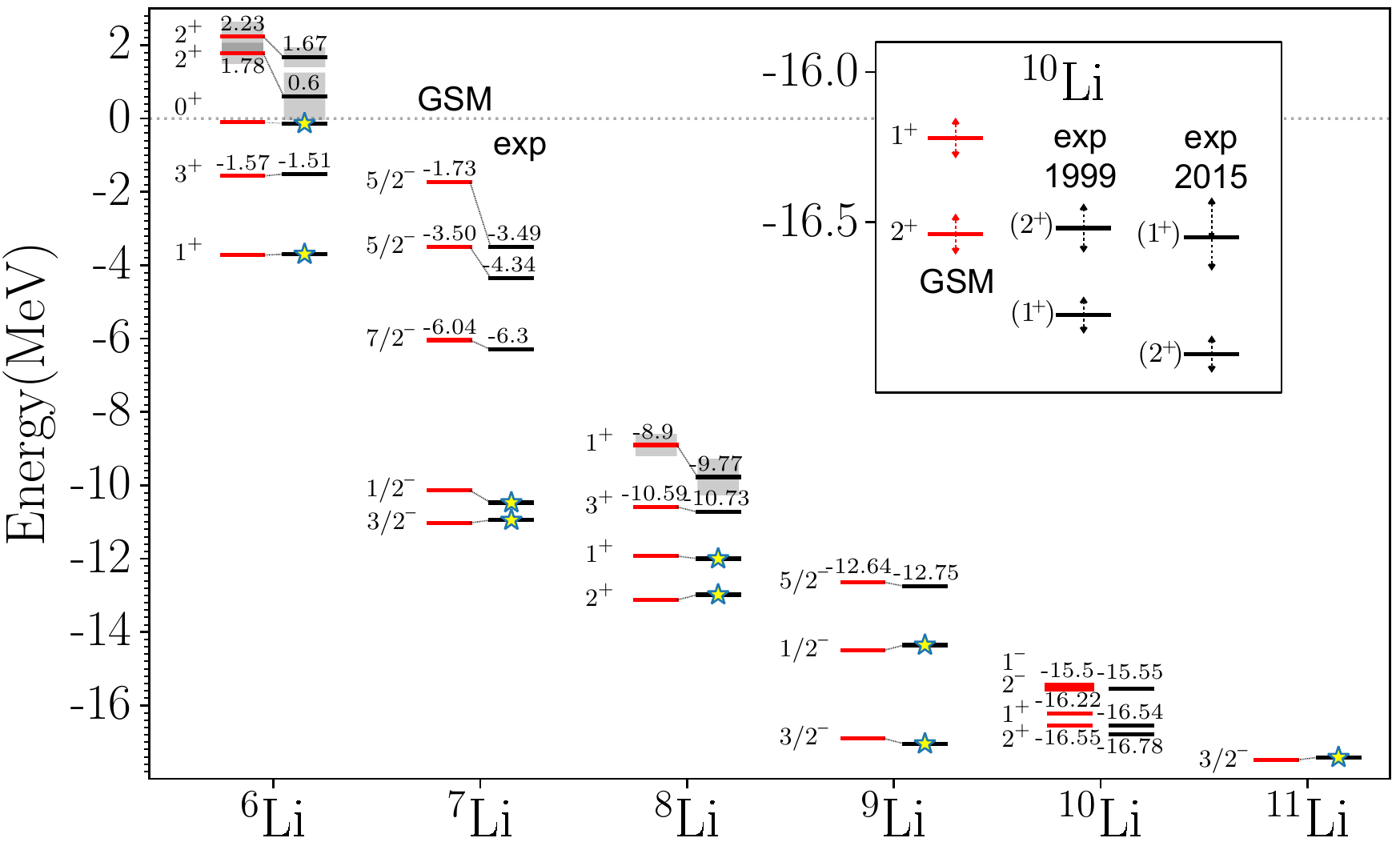}
\caption{\label{Fig_Li_isos} Level schemes of $^{6-11}$Li calculated in GSM and compared to experiment. 
Energies are given with respect to $^4$He core.
The resonance widths are marked by 
shaded boxes. The levels used in the GSM Hamiltonian optimization are marked by stars;
their energies are listed in Table~\ref{Table:Opt_states}.
Theoretical uncertainties for states not entering the optimization are given in Table.~\ref{Table:Opt_predict_states}.
The inset shows the predicted levels of $^{10}$Li compared to experimental data from 1999 \cite{bohlen_1999_1998} and 2015 \cite{smith_2015_1999}. Uncertainties on these levels are marked by arrows. See text for more discussion.
 }
\end{figure*}

As one can see in Table~\ref{Table:Opt_states}, a very good consistency between theoretical and experimental energies
has been achieved. The  root-mean-square  deviation from experimental values is  160\,keV. The largest discrepancy is obtained 
for the 1/2$^-$ states of $^7$Li and $^7$Be, where the deviation from the data is $\sim$340 keV.

\begin{table}[htb]
\caption{\label{Table.WSparam} Central and spin-orbit strengths  of the core-nucleon WS potential optimized in this work. The  statistical uncertainties are given in parentheses.}
\begin{ruledtabular}
\begin{tabular}{lcc}
Parameter & Neutrons & Protons \\
\hline\\[-7pt]
$V_0$ (MeV) & 39.5 (2) & 42.1 (4)\\
$V_{\ell\;\!\!s}$ (MeV\,fm$^2$) & 10.7 (2) &11.1 (5)
\end{tabular}
\end{ruledtabular}
\end{table}
The values of the parameters for the  WS potentials and the two-body interaction are displayed, along with their statistical uncertainties, in Tables~ \ref{Table.WSparam} and \ref{Table.two_body_int}, respectively.  As one can judge from the small parameter  uncertainties in  Tables~ \ref{Table.WSparam} and \ref{Table.two_body_int},
the GSM Hamiltonian fit is well constrained. As expected \cite{Yannen2017_1988}, the central term
$V_{c}^{00}$ has the largest uncertainty of $\sim$12\%.

\begin{table}[htb]
\caption{\label{Table.two_body_int}  Strengths  $V_\eta^{ST}$ of the two-body interaction 
optimized in this work. The  statistical uncertainties are given in parentheses.}
\begin{ruledtabular}
\begin{tabular}{l l l l}
$V_c^{01}$(MeV) 			&	&$-$9.425 (70)&	\\
$V_c^{10}$(MeV) 			&	&$-$8.309 (90)&	\\
$V_c^{00}$(MeV) 			&	&$-$8.895 (1130)&	\\
$V_T^{10}$(MeV\,fm$^{-2}$)	&	&$-$22.418 (970)&	
\end{tabular}
\end{ruledtabular}
\end{table}

It is to be noted that the  core-nucleon potential developed in the present study, optimized simultaneously with the two-body interaction,  is slightly shallower than the WS field
optimized in Ref.~\cite{Yannen2017_1988} to the
experimental $s$ and $p$ nucleon-$^4$He scattering phase
shifts. To assess the quality of the  WS potential obtained in this work, 
 Table~\ref{Table_He5_Li5} shows the predicted energies and  widths of the 3/2$^-$ g.s. 
of $^5$He and $^5$Li.
These values are indeed very close to predictions of Ref.~\cite{Yannen2017_1988} for 
$^5$He and $^5$Li.
\begin{table}[htb]
\caption{\label{Table_He5_Li5} Ground-state energies (in MeV) and widths (in keV)  of $^5$He and $^5$Li
obtained from the optimized core-nucleon  potential and compared to experiment \cite{tilley02_1931,tunl}.}
\begin{ruledtabular}
\begin{tabular}{c c c c c}
Nucleus	&	$E_{\rm GSM}$	&	 $E_{exp}$	& 	$\Gamma_{\rm GSM}$	& 	$\Gamma_{exp}$\\
\hline\\[-7pt]
$^5$He	&	0.74	&	 0.798		&	640		&	648 \\
$^5$Li	&	1.6	&	 1.69		&	1300		&	1230 \\
\end{tabular}
\end{ruledtabular}
\end{table}

Figure \ref{Fig_Li_isos} shows the energies calculated in the GSM  for the ground states and selected excited states in lithium isotopes.
Table~ \ref {Table:Opt_predict_states} lists the energy levels for states not entering the optimization with the corresponding statistical uncertainties.
As one can see, the optimized interaction allows for a good reproduction of experimental energies.
It is to be noted that the results for higher-excited states not included in the fit are also  very satisfactory.
For instance, the calculated $3^+$ state in $^6$Li at $-$1.57 MeV is only 60 keV below the experimental energy.
The experimental widths for the second $5/2^-$ state in $^7$Li (89\,keV) and $5/2^-$ state in $^9$Li (88\,keV) are very reasonable:
the GSM values are, respectively, 22 keV and 62 keV.
In general,  we do not expect the same quality of data reproduction for all excited states
due to the fact that the higher partial waves with $\ell \geq 2$, which may contribute to the wave functions of these states,
are not included in the model space.
The estimated statistical uncertainties on the predicted energies are small: in most cases they are in the range of 20-60\,keV.

\begin{table}[htb]
\caption{\label{Table:Opt_predict_states} Energy levels for states not entering the optimization.
The experimental values $E_{\text{exp}}$ are taken from \cite{ensdf}. The GSM values $E_{\text{GSM}}$ are shown with the uncertainties in the parenthesis.
}
\begin{ruledtabular}
\begin{tabular}{ c  c  c  c  }
	Nucleus	 &	State&	$E_{\text{exp}}$\,(MeV)		& 	$E_{\text{GSM}}$\,(MeV)		\\
\hline  \\[-6pt]
	$^6$Li 			&	3$^+$		& $-$1.51		& $-$1.57(2)					\\
	$^7$Li 			&	7/2$^-$ 	& $-$6.3		&$-$6.04(2)					\\
	$^8$Li			&	3$^+$ 	& $-$10.73		&$-$10.59(2)					\\
	$^9$Li 			&	5/2$^-$ 	& $-$12.75		&$-$12.64(2)				\\
\multirow{2}{*}{$^{10}$Li} &	2$^+$ 	& $-$16.78		&$-$16.55(5)  			\\
					&	1$^+$ 	& $-$16.54		&$-$16.22(5)				\\[3pt]
	$^7$Be 			&	7/2$^-$ 	& $-$4.73		&$-$4.47(2)					\\
	$^8$B 		 	&	3$^+$ 	& $-$7.12		&$-$7.11(2)				\\
	$^9$C  			&	5/2$^-$ 	& $-$7.14		&$-$7.12(5)						\\
\multirow{2}{*}{$^{10}$N} &	1$^-$ 	& $-$8.84		&$-$8.93(6)  			\\
					&	2$^-$ 	& $-$7.94		&$-$8.46(6)				
\end{tabular}
\end{ruledtabular}
\end{table}

\subsection{Structure of $^{10}$Li}\label{Li10_11}            

Several experiments \cite{Zinser_1995_2036,Thoennessen_1999_2035,Jeppesen_2006_2039,Simon_2007_2037}
and  theoretical studies \cite{Thompson_1994_2055,betan04_37} have 
indicated that the structure of the ground state in $^{10}$Li may correspond to  a valence neutron in a virtual $s$-state.
In a recent experiment \cite{Cavallaro_2017_1997}, the presence of an appreciable low-energy $\ell=0$  strength has not been confirmed.  Their conclusion was, however,  challenged in theoretical  studies  \cite{Moro_2019_2038,Barranco_2018_2040}.

\begin{figure*}[!htb]
\includegraphics[width=0.7\linewidth]{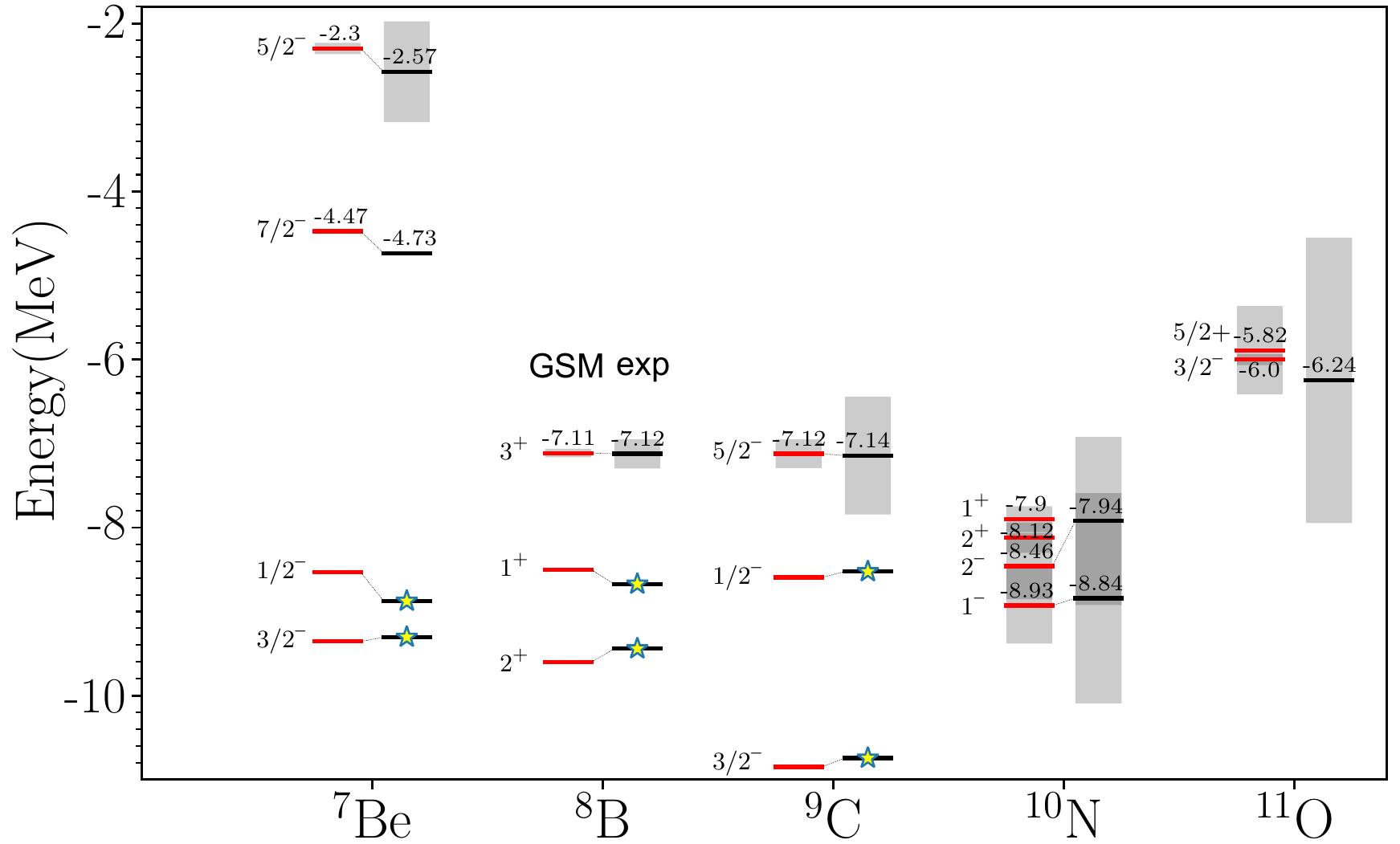}
\caption{\label{Fig_mirror_Li_isos} Similar to Fig.\ref{Fig_Li_isos} but for the mirror partners of the Li isotopes. 
Experimental energy of the $5/2^-$ resonance in $^9$C was taken from Ref.~\cite{Rogachev_1997_2007} and the data for $^{11}$O from Ref.~\cite{Webb_2019_2009}.}
\end{figure*}

We wish to note, however,  that a virtual state in  $^{10}$Li cannot be associated with an energy level of the system; the appearance of such a state in the complex-momentum plane manifests itself through a low-energy enhancement of the $n+^9$Li cross section, see Refs.~\cite{betan04_37,michel06_16,WangSM_2019_2008} for more discussion of this point in the context of the GSM description of $^{10,11}$Li.
For that reason, we limited our calculations to resonant states in $^{10}$Li that can be interpreted as experimentally-observable resonances.

The computed ground-state 2$^+$ and the first excited state 1$^+$ are  predicted, respectively,   at 0.35\,MeV and 0.68\,MeV above the $n+^9$Li threshold. As seen in 
Fig.\,\ref{Fig_Li_isos},
the practically degenerate  1$^-$  and 2$^-$ states are calculated at 1.05\,MeV.
A comment is in order here. To achieve the numerical stability, the calculation of the resonances in $^{10}$Li
had to be performed by employing a basis-generating WS potential that is deeper than the optimized core-nucleon potential.
We have checked that in this way we could obtain very stable results for  the energies, with accuracy below 1\,keV. On the other hand, the computed widths, of the order of few hundreds keV,  are  not stable. For that reason, we do not show them in Fig.~\ref{Fig_Li_isos}.

Table~\ref{Table.Li10_confi} lists the squared amplitudes of the dominant neutron configurations for the four low-lying states of $^{10}$Li. The positive parity states 
 are  primarily  made from the $0p_{3/2}$ and $0p_{1/2}$ resonant shells.
The negative parity states contain one neutron in the $1s_{1/2}$  shell. The contribution from the non-resonant continuum space to the low-lying states is very small.

\begin{table}[htb]
\caption{\label{Table.Li10_confi}  Squared amplitudes of dominant configuration of valence neutrons and protons for low-lying levels of $^{10}$Li and $^{10}$N, respectively.
The odd proton in $^{10}$Li and the odd neutron in $^{10}$N occupy the  $0p_{3/2}$ Gamow state.
The tilde sign labels  non-resonant continuum  components.}
\begin{ruledtabular}
\begin{tabular}{c |c c |c c}
configuration 	&	\multicolumn{2}{c|}{$^{10}$Li}&	\multicolumn{2}{c}{$^{10}$N}	\\
\hline\\[-7pt]
								&	$2^+$		&	$1^+$	&	$2^+$	&$1^+$	\\
$ (0p_{3/2})^4 (0p_{1/2})^1$				& 0.84	&	0.81	&	0.81	&	0.78\\
$ (0p_{3/2})^3 (0p_{1/2})^2$				& 0.10	&	0.06 &	0.10	&	0.05\\
\hline\\[-7pt]
	 							&	$1^-$	&	$2^-$	&$1^-$	&$2^-$	\\
$ (0p_{3/2})^4 (1s_{1/2})^1$				& 0.72	&	0.73	&	0.44	&	0.37	\\
$ (0p_{3/2})^4 (\widetilde{s_{1/2}})^1$		&		&		&	0.29	&	0.35	\\
$ (0p_{3/2})^3 (0p_{1/2})^1 (1s_{1/2})^1$		& 0.14	&	0.14	&	0.09	&	0.07	\\
$(0p_{3/2})^3  (0p_{1/2})^1 (\widetilde{s_{1/2}})^1$	&	&		&	0.06	&	0.07\\
$ (0p_{3/2})^2 (0p_{1/2})^2 (1s_{1/2})^1$		& 0.07	&	0.07	&	0.04	&	0.03	\\
$(0p_{3/2})^2 (0p_{1/2})^2 (\widetilde{s_{1/2}})^1$	&	&		&	0.03	&	0.03	
\end{tabular}
\end{ruledtabular}
\end{table}

In Ref.~\cite{bohlen_1999_1998} they
observed two positive-parity states at 0.24\,MeV and 0.53\,MeV above the $n+ ^9$Li threshold. The $J^\pi=1^+$  assignment for the lower state   was questioned in 
 Ref.~\cite{smith_2015_1999} who suggested a $J^\pi=2^+$ assignment, see the inset in Fig.~\ref{Fig_Li_isos}. Considering the large experimental widths of the 
$1^+/2^+$ doublet, 0.10/0.4\,MeV ~\cite{bohlen_1999_1998} or
 0.8/0.2\,MeV \cite{smith_2015_1999}, both experimental results are consistent with the GSM  results.
The computed position of the negative-parity $1^-, 2^-$ doublet is consistent with the observation of a negative-parity state at $\sim$1.5\,MeV \cite{Cavallaro_2017_1997}.

\subsection{Mirror partners of lithium isotopes} {\label{Mirror_nuclei}}

The level schemes  for the mirror partners of lithium isotopes are shown in Fig.~\ref{Fig_mirror_Li_isos}.
As in the Li case, we obtain a very reasonable  agreement with experiment.
The $5/2^-$ and $7/2^-$ excited states in $^7$Be are slightly ($<300$keV) above the corresponding experimental values, 
whereas the position of the resonant $3^+$ states in  $^8$B and  $5/2^-$ state in $^9$C
are well reproduced, as well as the 
weakly-bound g.s. of  $^8$B and $^9$C.

In the following  we focus on the unbound nuclei $^{10}$N and $^{11}$O.
Due to the presence of the Coulomb barrier, the $1s_{1/2}$ single-proton state  is a resonance rather than a virtual state \cite{WangSM_2019_2008,Webb_2019_2009}. To capture this state,
a complex contour is  used with  a $k_{\rm peak}=(0.25-0.05i)$\,fm$^{-1}$.

The spectrum of $^{10}$N is not experimentally known with certainty.
In Fig.~\ref{Fig_mirror_Li_isos}, we show the tentative level assignments used in Ref.~\cite{ensdf}.  According to Refs.~\cite{Sherr_2013_2029,fortune13_1910}, the ground state of $^{10}$N  is most likely a $1^-$ state of energy in the range from 1.81 to 1.94\,MeV. In a more recent work \cite{Hooker_2017_2026}, they observed two low-lying negative-parity states  but they were not able to assign $J^\pi$ values.

Our calculations for $^{10}$N  predict
the ground state to be a $1^-$ state with  
($E, \Gamma$)=($-$8.93,0.9)\,MeV that  lies 1.92 MeV above the one-proton threshold. The first excited state is predicted to be a 2$^-$ state  with $\Gamma$=0.3\,MeV  slightly below the  value quoted in  Ref.~\cite{Hooker_2017_2026}. This result is consistent with the recent Gamow coupled-channel analysis of Ref.~\cite{WangSM_2019_2008}. 
We also predict an excited  1$^+$ state with $\Gamma=0.3$\,MeV, lying 2.9 MeV above the $^9$C+$p$ threshold, as well as  a second positive-parity  $2^+$ state  with a width of 0.36\,MeV.

Table~\ref{Table.Li10_confi} shows the squared amplitudes of the dominant proton configurations for the four low-lying states of $^{10}$N. Similar to $^{10}$Li, the positive parity states 
 are  primarily  made from the $0p_{3/2}$ and $0p_{1/2}$ resonant shells.
The dominant configurations of negative parity states contain one $\ell=0$ proton, which can  either be in   the $1s_{1/2}$  shell or in a non-resonant continuum state.

The unbound $^{11}$O is the mirror partner of the $2n$-halo nucleus $^{11}$Li.
The first observation of $^{11}$O was achieved recently \cite{Webb_2019_2009}. A broad peak with a  width of $~3.4$ MeV was observed which was interpreted in terms of four overlapping $3/2^-$ and $5/2^+$ resonances.
Our GSM calculations predict a $3/2_1^-$ g.s. with a width of 0.13\,MeV and 
the first excited  $5/2_1^+$ state  with $\Gamma\approx 1$\,MeV, see  Fig.~\ref{Fig_mirror_Li_isos}.
These predictions are consistent with  the Gamow coupled-channel calculations of Ref.~\cite{WangSM_2019_2008}.

\begin{figure}[!htb]
\includegraphics[width=1.0\linewidth]{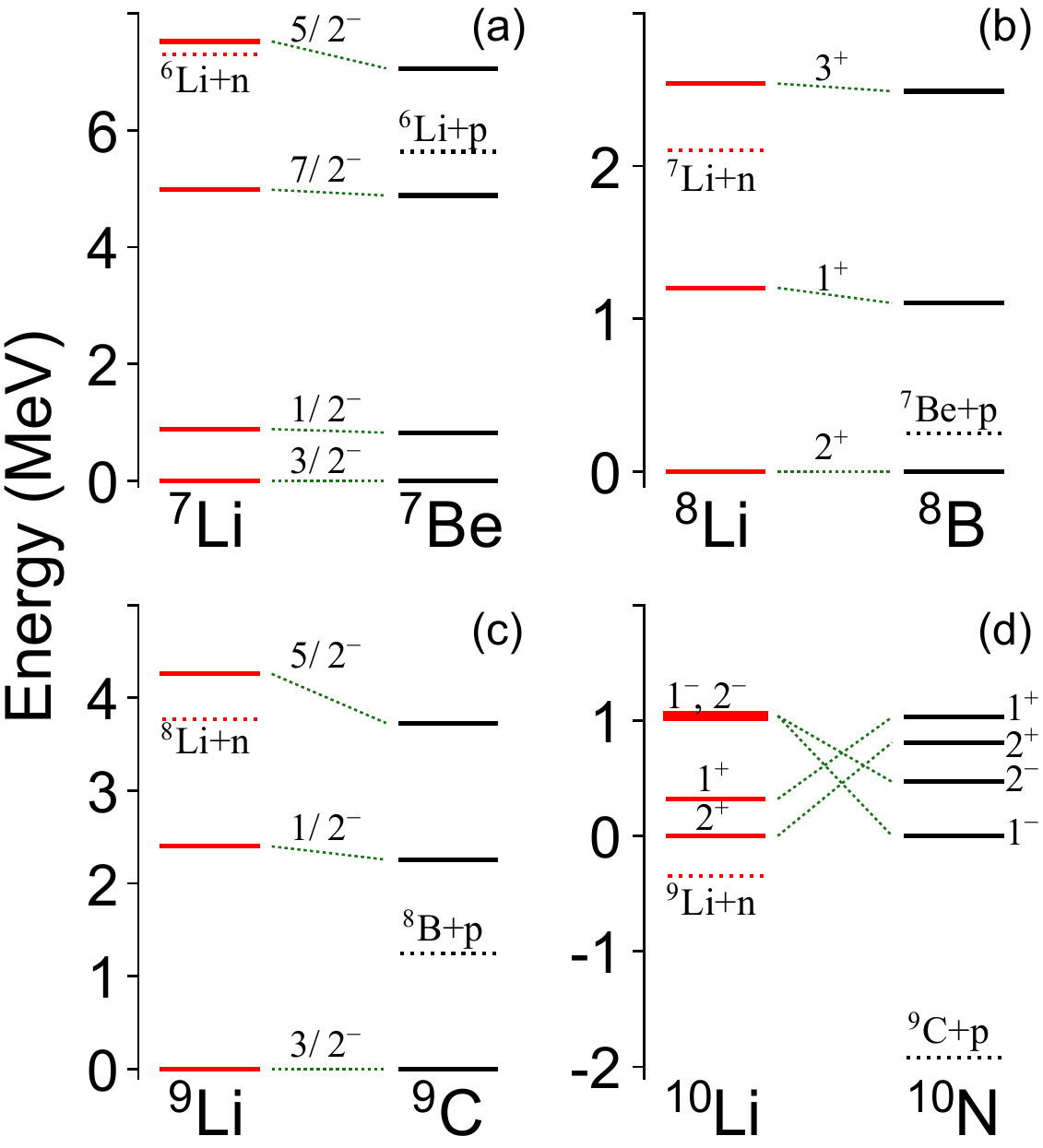}
\caption{\label{Fig_ThomasE}
Level schemes of  Li isotopes with (a) $A=7$, (b) $A=8$, (c) $A=9$, (d) $A=10$,  and their mirror partner predicted in our GSM calculations. The energies are plotted with respect to the g.s. energy (at zero). The one-nucleon  emission thresholds are  marked.}
\end{figure}

To study the effect of particle continuum due to  different positions of particle thresholds  in  mirror partners, or Thomas-Ehrman effect  \cite{ehrman51_592,thomas52_593}, in Fig.~\ref{Fig_ThomasE} we compare the level schemes of  Li isotopes and their mirror partners. 
(For the early GSM study of the Thomas-Ehrman shifts in light nuclei, see Ref.~\cite{Michel_2019_2063}.)
As expected, the proton-unbound states in proton-rich mirror nuclei are shifted down in energy as compared to the states in neutron-rich partners, which lie below, or slightly above the one-neutron threshold.

The $^{10}$Li-$^{10}$N mirror pair is the most interesting one as both nuclei lie above the  particle-emission thresholds. As seen in 
Table~\ref{Table.Li10_confi}, the effect of the very low $^9$C+$p$ threshold in $^{10}$N
on  the negative-parity states 1$^-$ and 2$^-$  containing the  $s$-wave  proton is huge:
it results in a rather dramatic shift of both negative parity states
 when
going from  $^{10}$Li to $^{10}$N 
that  gives rise to a different structure of low-lying resonances in these nuclei.

\section{Conclusions}\label{conclusion}

In this work, we studied level schemes of $^{6-11}$Li  and their  mirror partners in the framework of the complex-energy Gamow shell model assuming the rigid $^4$He core.  The effective interaction
between valence nucleons is constructed based on a simplified version of the FHT potential.

By fitting four FHT coupling constants and four parameters of the core-nucleon potential,
to  the experimental energies of 15 states in $^{6-9,11}$Li, $^7$Be, $^8$B and $^9$C, we managed to
construct a well constrained interaction. A rms deviation from experiment of 160\,keV was reached for energy levels used in the GSM Hamiltonian optimization, with the statistical errors of the GSM  Hamiltonian parameters  not exceeding 12\%.
This result suggests that the ``complex-made-simple" scenario proposed in Ref.~\cite{Fossez_2018_1994} for the He chain also works for heavier nuclei involving valence protons. Namely,  a parameter reduction guided by effective-scale arguments provides a practical alternative to full-fledged $A$-body calculations for drip-line nuclei. 

We assessed the predictive power of the optimized  Hamiltonian by making predictions for  excited states not included in the fit. In general, a very reasonable agreement with testing data was obtained, see Table~\ref{Table:Opt_predict_states}.

Predictions were also made for the particle-unstable nuclei $^{10}$Li, $^{10}$N, and $^{11}$O. The computed  3/2$^-$  ground state  of  $^{11}$O  is consistent with the recent   Gamow coupled-channel calculations \cite{Webb_2019_2009,WangSM_2019_2008}.
The ground state of $^{10}$Li is predicted to be a 2$^+$ state about 0.35 MeV above the neutron-emission  threshold, in accordance with Ref.~\cite{smith_2015_1999}
while the lowest negative-parity state 1$^-$ is expected to lie $\sim$1.0\,MeV higher, in agreement with Ref.~\cite{Cavallaro_2017_1997}.
Due to a spectacularly strong Thomas-Ehrman effect, for  $^{10}$N we predict  the $1^-$ ground state  and $2^-$ first excited state.

By successfully reproducing the structure of lithium isotopes and their mirror partners  with an optimized  interaction, we demonstrated that the quantified GSM is capable of quality predictions for  exotic light nuclei with several valence protons and neutrons.
Our  future efforts will focus on Be and B isotopes, 
which exhibit complex structure due to the intricate effects of continuum coupling and clustering \cite{pillet07_1879,LASHKO_2017_2072,DELLAROCCA_2018_2074,GARRIDO_2010_2073,Alvarez_2010_2075}.


\begin{acknowledgments}
Discussions with Kevin Fossez, Marek P{\l}oszajczak, and Simin Wang are gratefully acknowledged as well as comments from Joshua Wylie. Computational resources were provided by the Institute for Cyber-Enabled Research at Michigan State University. This material is based upon work supported by the U.S.\ Department of Energy, Office of Science, Office of Nuclear Physics under  award numbers DE-SC0013365 (Michigan State University) and  DE-SC0018083 (NUCLEI SciDAC-4 collaboration), and by the International Scientific Cooperation Conicet-NSF 1225-17.
\end{acknowledgments}

%

\end{document}